\documentclass[prd,aps,twocolumn,amsmath,amssymb,nofootinbib,preprintnumbers]
{revtex4}

\usepackage{graphicx}
\usepackage{dcolumn}
\usepackage{bm}
\usepackage{amsmath}
\usepackage{amsfonts}
\usepackage{euscript,bbm}
\usepackage{ifthen}
\usepackage{psfrag}
\usepackage{slashed}

\def\ls{\mathrel{\lower4pt\vbox{\lineskip=0pt\baselineskip=0pt
           \hbox{$<$}\hbox{$\sim$}}}}
\def\gs{\mathrel{\lower4pt\vbox{\lineskip=0pt\baselineskip=0pt
           \hbox{$>$}\hbox{$\sim$}}}}
\def\drawbox#1#2{\hrule height#2pt

\hbox{\vrule width#2pt height#1pt \kern#1pt
              \vrule width#2pt}
              \hrule height#2pt}

\def\Asym#1#2{\vcenter{\vbox{\drawbox{#1}{#2}
              \kern-#2pt       
              \drawbox{#1}{#2}}}}

\newcommand{\be}{\begin{equation}}
\newcommand{\ee}{\end{equation}}
\newcommand{\bea}{\begin{eqnarray}}
\newcommand{\eea}{\end{eqnarray}}

\begin{document}

\title{A Supersymmetric Model for Dark Matter and Baryogenesis Motivated by the Recent CDMS Result}

\author{Rouzbeh Allahverdi$^{1}$}
\author{Bhaskar Dutta$^{2}$}
\author{Rabindra N. Mohapatra$^{3}$}
\author{Kuver Sinha$^{2}$}

\affiliation{$^{1}$~Department of Physics and Astronomy, University of New Mexico, Albuquerque, NM 87131, USA \\
$^{2}$~Mitchell Institute of Fundamental Physics and Astronomy, Department of Physics and Astronomy, Texas A\&M University, College Station, TX 77843-4242, USA\\
$^{3}$~Maryland Center for Fundamental Physics, Department of Physics, University of Maryland, College Park, MD 20742, USA}

\begin{abstract}
We discuss a supersymmetric model for cogenesis of dark and baryonic matter where the dark matter (DM) has mass in the 8-10 GeV range as indicated by several direct detection searches including most recently the CDMS experiment with the desired cross section. The DM candidate  is a real scalar filed.  Two key distinguishing features of the model are the following: (i) in contrast with the conventional WIMP dark matter scenarios where thermal freeze-out is responsible for the observed relic density, our model uses non-thermal production of  dark matter after reheating of the universe caused by moduli decay at temperatures below the QCD phase transition, a feature which alleviates the relic over-abundance problem caused by small annihilation cross section  of light DM particles; (ii) baryogenesis occurs also at similar low temperatures from the decay of TeV scale mediator particles arising from moduli decay. A possible test of this model is the existence of colored particles with TeV masses accessible at the LHC.
\end{abstract}
MIFPA-13-14 \\  April, 2013
\maketitle

\noindent
{\it {\bf Introduction-}} The CDMS Collaboration \cite{Agnese:2013dwa} has recently announced results from a blind analysis of data taken with Silicon detectors of the CDMSII experiment in 2006-2007. The collaboration reports dark matter (DM) events that survive cuts with a significance of $3.1 \sigma$ corresponding to DM mass $m_{\rm DM} \sim 8$ GeV and spin-independent scattering cross-section $\sigma_{\rm SI} \sim 10^{-41}$ cm$^{2}$. The excess reported by the CoGeNT collaboration \cite{cogent} hints at light dark matter in a similar region of parameter space, while CDMS II Ge \cite{cdmsGe} and EDELWEISS \cite{edelweiss} data do not exclude it. While XENON100 data \cite{xe100} would appear to rule out this result at the present time, XENON10 \cite{xe10data} is not that inconsistent with it \cite{xe10}, clearly warranting further probes of this region.


If a light dark matter with cross sections given above is confirmed, it will pose a challenge to most scenarios where DM is a weakly interacting massive particle (WIMP)\cite{feng} e.g., the conventional ones in the context of the minimal supersymmetric standard model (MSSM), since exchange of ${\cal O}$(TeV) particles would lead to a smaller cross section for such low masses and hence an over-abundance of relic DM at the current epoch assuming standard cosmological evolution. On the other hand, this is suggestive of scenarios which address the DM-baryon asymmetry coincidence problem, that focus on the curious observation that the energy densities in baryons and DM are of the same order of magnitude (roughly $\sim 1:5$ \cite{WMAP}) often despite the quite different mechanisms used to generate them \cite{Coincidence}. It
would seem natural to point towards scenarios in which this apparent coincidence is addressed by an underlying connection between the DM production and baryogenesis scenarios, such that the \textit{number densities of DM and baryons are roughly equal}.

In this work, we present a simple extension of MSSM which has a DM candidate of ${\cal O}(10~{\rm GeV})$ mass and a desired scattering cross-section resulting from the exchange of a new TeV scale colored particle. It also implements a low-scale baryogenesis scenario without adding any extra features and addresses the coincidence problem. Satisfying the DM scattering cross-section typically leads to a region of parameter space where thermal freeze-out gives an over-abundance of DM particles. We thus rely on non-thermal DM production \cite{MR} which, in this context, is useful in several ways: $(i)$ the over-abundance of thermal DM can be addressed within a non-thermal scenario by producing the correct number density from a late decay without relying on further DM annihilation \cite{Visible}, $(ii)$ non-thermal baryogenesis can be achieved with $\mathcal{O}(1)$ couplings of the new fields to the MSSM fields \cite{Late} and $(iii)$ the coincidence problem is addressed through the framework of Cladogenesis \cite{Clado}, in which the dilution factor due to the decay of a modulus field is mainly responsible for the observed relic densities, while roughly equal number densities for baryons and DM may be obtained due to comparable branching fractions of the DM and the baryon asymmetry per modulus decay.

We emphasize that the DM candidate in our model is a scalar field, which is needed in order to generate a large DM-nucleon cross-section hinted by the recent CDMSII results. Recently, in an attempt to explain the coincidence problem, we showed \cite{ad} that the non-supersymmetric version of the model can naturally give rise to a fermionic DM candidate with a mass on the order of the proton mass. However, $\sigma_{\rm SI}$ is hierarchically smaller in this case due to the Majorana nature of the DM candidate. The difference between the two scenarios may be also distinguished at colliders.

\noindent
{\it {\bf The Model-}} We start with the MSSM and introduce new iso-singlet color-triplet superfields $X$ and ${\bar X}$ with respective hypercharges $+4/3$ and $-4/3$, and a singlet superfield $N$ with the following superpotential \cite{Rabi2}
\bea \label{superpot}
W & = & W_{\rm MSSM} + W_{\rm new} \, \nonumber \\
W_{\rm new} & = & \lambda_{i} X N u^c_{i} + \lambda^\prime_{i j} {\bar X} {d^c_i} d^c_j + M_X X {\bar X} + \frac{M_N}{2} N N \, . \nonumber \\
& & \,
\eea
Here $i,~j$ denote flavor indices (color indices are omitted for simplicity), with $\lambda^\prime_{i j}$ being antisymmetric under $i \leftrightarrow j$. We assume the new colored particles associated with the $X,~\bar{X}$ superfields to have TeV to sub-TeV mass and the scalar partner of singlet $N$, denoted by $\tilde{N}$, will be assumed to have mass in the $8-10$ GeV range and will be identified with the DM particle. We first wish to clarify that even though the particle content and the superpotential of this model is identical to that in Ref. \cite{Rabi2}, the cosmological scenario outlined here is vastly different, as we describe below.

There are already constraints on the parameters of the model from observations for the assumed mass range of the particles above. The exchange of $X,\bar{X}$ particles in combination with the Majorana mass of $N$ lead to $\Delta B=2$ and $\Delta S=2$ process of double proton decay $pp \to K^+K^+$. Current experimental limits on this process from the Super-Kamiokande experiment \cite{miura} imply that the combination $\lambda^2_1 \lambda^2_{12} \leq 10^{-10}$ for $M_N\sim 100$ GeV. Since we will need $\lambda_i \sim 1$ for further considerations, the above constraint implies that $\lambda_{12} \sim 10^{-5}$. We also note that $M_N \gs {\cal O}({\rm GeV})$ is needed in order to avoid rapid proton decay $p \rightarrow N + e^+ + \nu_e$ (if $M_N \approx m_p$, the fermionic component of $N$ can be the DM candidate but $\sigma_{\rm SI}$ will be much smaller than that indicated by the CDMS experiment \cite{ad}).

To discuss the DM candidate ${\tilde N}$, we note that after supersymmetry breaking, the real and imaginary parts of this field acquire different masses
\be \label{mass}
m^2_{{\tilde N}_{1,2}} = M^2_N + m^2_{\tilde N} \mp B_N M_N ,
\ee
where $m_{\tilde N}$ is the soft breaking mass of ${\tilde N}$ and $B_N M_N$ is the $B$-term associated with the $M_N N^2/2$ term in the superpotential. We have assumed that $B_N M_N$ is positive, which can be achieved by a proper field rotation. The lighter of the two mass eigenstates ${\tilde N}_1$ will be assumed to be the lightest supersymmetric particle (LSP). We assign quantum number $+1$ under $R$-parity to the scalar components of $X,~{\bar X}$ and the fermionic components of $N$. The scalar component of the $N$-superfield will then have odd R-parity. $R$-parity conservation then guarantees the stability of the LSP, $\tilde{N_1}$, which then becomes the DM candidate.
One can make ${\tilde N}_1$ arbitrarily light by adjusting the three terms in Eq.~(\ref{mass}) that contribute to $m_{{\tilde N}_1}$. If $M_N \sim m_{\tilde N} \sim B \sim M$, the level of tuning needed to get $m_{{\tilde N}_1} \ll M$ will be $\delta \sim m_{{\tilde N}_1}/M$. For example, if $M \sim {\cal O}(100 ~ {\rm GeV})$, tuning at the level of $10\%$ is needed in order to have $m_{{\tilde N}_1} \sim {\cal O}(10 ~ {\rm GeV})$.

The superpotential coupling $\lambda_i X N u^c_i$ yields an effective interaction between ${\tilde N}_1$ and a quark $\psi$ via $s$-channel exchange of the fermionic component of $X$. The amplitude is given by $i \frac{\vert \lambda_1 \vert ^2}{4M^2_X}({\bar \psi}(k^{\prime}) \gamma^\mu \psi(k))Q_\mu $, where $k_\mu$ is the quark momentum, $p_\mu$ is the momentum of ${\tilde N}_1$, and $Q_\mu = k_\mu+p_\mu$.
%
%
This results in the following spin-independent DM-proton elastic scattering cross section
\be \label{SI}
\sigma^{\rm SI}_{{\tilde N}_1-p} \simeq \frac{\vert \lambda_1 \vert^4}{16 \pi} \frac{m^2_p}{M^4_X},
\ee
where $m_p$ is the proton mass \cite{Rabi2}. It is seen that for $\vert \lambda_1 \vert \sim 1$ and $M_X \sim 1$ TeV, which is compatible with the LHC bounds on new colored fields \cite{LHCcolored}, we get $\sigma^{\rm SI}_{{\tilde N}_1-p} \sim {\cal O}(10^{-41})$ cm$^2$. We note that this scenario easily evades bounds coming from monojet searches at colliders \cite{Cheung:2012gi}. The pair production of fermionic components of $X,~{\bar X}$ superfields, which are $R$-parity odd, will produce 4 jets plus missing energy final states at the LHC in this model. In the non-supersymmetric version of the model \cite{ad}, where $N$ fermion is the DM candidate, the absence of $R$-parity fields results in  missing energy final states with 2 and 3 jets only, which will allow us to distinguish the two scenarios.
\vskip 2mm
\noindent
{\it {\bf Dark Matter Production and Baryogenesis-}} The superpotential coupling $\lambda_i X N u^c_i$ also results in annihilation of ${\tilde N}_1$ quanta into a pair of a right-handed quark and left-handed antiquark of the up-type. Considering that $m_{{\tilde N}_1} \sim {\cal O}(10~{\rm GeV})$, only annihilation to up and charm quarks is possible when temperature of the universe is below $m_{{\tilde N}_1}$. The annihilation rate is given by
\be \label{ann}
\langle \sigma_{\rm ann} v_{\rm rel} \rangle \simeq \frac{\vert \lambda_1 \vert^4 + \vert \lambda_2 \vert^4 +2\vert\lambda_1\lambda^*_2\vert^2}{8 \pi} \frac{\vert {\vec p} \vert^2}{M^4_X} ,
\ee
where ${\vec p}$ is the momentum of annihilating ${\tilde N}_1$ particles. It is seen that for $\vert \lambda_1 \vert \sim \vert \lambda_2 \vert \sim 1$, $m_{{\tilde N}_1} \sim {\cal O}(10 ~ {\rm GeV})$, $M_X \sim 1$ TeV we have $\langle \sigma_{\rm ann} v_{\rm rel} \rangle_{\rm thermal} \ll 3 \times 10^{-26}$ cm$^3$ s$^{-1}$. Therefore thermal freeze-out yields an over-abundance of ${\tilde N}_1$ particles.

This implies that obtaining the correct DM relic density requires a non-thermal scenario. An attractive scenario involves a scalar field $S$ whose late decay reheats the universe below the freeze-out temperature $T_{\rm f}$ of DM annihilation, dilutes the over-abundant relics to negligible levels via extra entropy production, and simultaneously produces DM particles \cite{kolb}. Such a scenario could arise naturally in string theory inspired models where $S$ could be a modulus with only gravitational couplings to the visible sector fields. Following the decay of $S$, two options are possible: $(i)$ DM particles produced from the decay of $S$ undergo further DM annihilation or $(ii)$ no further annihilation occurs. The first option can happen if $\langle \sigma_{\rm ann} v_{\rm rel} \rangle_{\rm thermal} > 3 \times 10^{-26}$ cm$^3$ s$^{-1}$, which implies thermal under-abundance of DM particles. However, this option is not available in our model since, as mentioned, thermal freeze-out yields an over-abundance of DM particles.

Thus an important requirement for implementing the late decay scenario in our model is that
the branching ratio for the production of $R$-parity odd particles (which eventually decay to ${\tilde N}_1$) from $S$ decay must have the correct magnitude to yield the right DM abundance directly. In this connection, it is worth noting that the super-partner of the modulus field which is also weakly coupled does not pose any challenge to cosmology as, if present after inflation, its energy density is subdominant to that of $S$ and decays along with it.

The field $S$ can either be a gravitationally coupled modulus \cite{MR,Higgsino} or a heavy scalar belonging to the visible sector \cite{Visible}. Here, we give a brief outline of the first option, keeping the detailed model-building for future work. In a plausible scenario, $S$ mainly decays into scalar components of $X,~{\bar X}$ superfields (denoted by ${\tilde X},~{\tilde {\bar X}}$ respectively), which are $R$-parity even fields. This can be achieved through a coupling $K \supset \lambda_X S^\dagger {\tilde X} {\tilde {\bar X}}$ in the K\"ahler potential.
%
The decay into the $R$-parity even fermions suffers chiral suppression. The decays of $S$ to $R$-parity odd gauginos can be suppressed by suitable geometric criteria e.g., by constructing the visible sector at a singularity and selecting $S$ to be the volume modulus in large volume compactification scenarios \cite{Cicoli:2012aq}. The decay of $S$ to other $R$-parity odd MSSM fields like squarks and sleptons is suppressed after using the equations of motion. The decay to the gravitino can also be suppressed for superheavy ($\sim 10^{12}$ GeV) gravitinos, which enables one to avoid overproduction of DM by late-time gravitino decay \cite{Gravitino}. Finally, the decay of $S$ to ${\tilde N}_{1,2}$ is suppressed by preventing the K\"ahler potential coupling $\lambda_N S^\dagger {\tilde N}^2$.

The above scenario can be achieved in a natural manner by considering the theory to be invariant under a discrete symmetry $Z_{18}$. The various fields have the following quantum numbers under $Z_{18}$ (which happens to be a subgroup of baryon number) given in the table below.

\begin{center}
Table 1. Charge assignments of the various fields under the discrete symmetry $Z_{18}$ ( fields not in the table are neutral).
\vskip 3mm
\begin{tabular}{|c||c|}\hline\hline
Fields & $Z_{18}$ transformation\\\hline
$(u^c,d^c),Q$ & $e^{\frac{-i\pi}{9}}(u^c,d^c),e^{\frac{i\pi}{9}}Q$\\
$(X,\bar{X})$ & $(e^{\frac{-2i\pi}{9}}X,e^{\frac{2i\pi}{9}}\bar{X})$\\
$N$ & $e^{\frac{i\pi}{3}}N$\\
$\delta$ & $e^{\frac{-2i\pi}{3}}\delta$\\\hline\hline
\end{tabular}
\end{center}
\vskip 3mm
The superpotential $W_{\rm new}$ in Eq. (1) is now replaced by:
\bea \label{superpot1}
W_{\rm new} & = & \lambda_{i} X N u^c_{i} + \lambda^\prime_{i j} {\bar X} {d^c_i} d^c_j + M_X X {\bar X} + f\delta N N +\kappa \delta^3\, .
\nonumber \\
& & \,
\eea
The scalar component of the $\delta$ superfield (also denoted by $\delta$) will be assumed to acquire a vacuum expectation value (VEV) after supersymmetry breaking to give rise to the mass term $M_N N^2/2$ in Eq. (1) with $M_N=2f<\delta>$.
One can have the following K\"ahler term
$K \supset \lambda_N S^\dagger \delta NN/M_{\rm P}$
where the modulus $S$ is a singlet under
the $Z_{18}$.
It is important to note that the coupling of $S$ to $\tilde{N}$ is suppressed $\propto \langle \delta \rangle/M_{\rm P}$ compared to its coupling to $\tilde{X} \tilde{\bar{X}}$, which arises without any Planck mass suppression. As a result, $S$ field will predominantly decay to $\tilde{X}\tilde{\bar{X}}$ rather than $\tilde{N}$ as assumed above.


DM and ordinary matter will be produced in subsequent decay of ${\tilde X}$ and ${\tilde {\bar X}}$. The abundance of DM particles thus produced is given by
\be \label{DMab}
\frac{n_{{\tilde N}_1}}{s} = Y_S {\rm Br}_{{\tilde N}_1}.
\ee
Here $Y_S \equiv 3 T_{\rm r}/4 m_S$ is the dilution factor due to $S$ decay, where $m_S$ and $T_{\rm r}$ are mass of $S$ and reheat temperature of the universe from $S$ decay respectively. ${\rm Br}_{{\tilde N}_1}$ denotes the branching ratio for producing $R$-parity odd particles from the decay of ${\tilde X},~{\tilde {\bar X}}$.

Assuming that the squarks and gluinos are heavier than ${\tilde X},~{\tilde {\bar X}}$, the decays of latter do not produce any $R$-parity odd particles at the leading order. Decay to $d^c_i d^c_j$ and $N u^c_i$ final states results in a decay width $\Gamma_{{\tilde X},{\tilde {\bar X}}} \sim (\vert \lambda_i \vert^2 + \vert \lambda^{\prime}_{ij} \vert^2) m_{\tilde X}/8 \pi$, where $m_{\tilde X}$ denotes the mass of ${\tilde X},~{\tilde {\bar X}}$. Three-body decays into $u^c_i {\tilde N} {\tilde B}$ can produce $R$-parity odd particles provided that the Bino ${\tilde B}$ is lighter than ${\tilde X},~{\tilde {\bar X}}$. This leads to ${\rm Br}_{{\tilde N}_1} \sim 10^{-3}$.

The measured DM relic abundance for $m_{{\tilde N}_1} \sim {\cal O}(10~{\rm GeV})$ is $(n_{{\tilde N}_1}/s) \approx 5 \times 10^{-11}$. One therefore needs a dilution factor $\sim 5 \times 10^{-8}$, which can be achieved for $m_S \sim 1000$ TeV and $T_{\rm r} \sim 10$ MeV. For a decay width $\Gamma_S = (c/2\pi) (m^3_S/M^2_P)$, the reheat temperature is given by $T_r \sim c^{1/2} (m_S/100 {\rm TeV})^{3/2} \times 10$ MeV. Thus, one requires $c \sim 0.01$, which can be obtained in specific constructions.

$S$ decay substantially dilutes any previously generated baryon asymmetry. Since $T_{\rm r} \sim 10$ MeV, a mechanism of post-sphaleron baryogenesis \cite{Rabi1} is required to produce the desired value of baryon asymmetry $\eta_B \sim 10^{-10}$, where $\eta_B \equiv (n_B - n_{\bar B})/s$.  The asymmetry will be generated from the $S$ decay dilution factor times the baryon asymmetry ($\epsilon$) generated from  the decay of ${\tilde X},~{\tilde {\bar X}}$. A minimal set up includes two copies of ${\tilde X},~{\tilde {\bar X}}$ fields and the interference between tree-level and one-loop self-energy diagrams gives rise to the baryon asymmetry. In Fig. 1, we show diagrams responsible for generating baryon asymmetry from ${\tilde X}_1,~{\tilde {\bar X}}_1$ decays. Since the the dilution factor is $10^{-8}$, we need $\epsilon \sim 10^{-2}$ and the masses of ${\tilde X}_1, {\tilde X}_2$ do not need to be close in our scenario.

\begin{figure}[ht]
\centering
\includegraphics[width=3.6in]{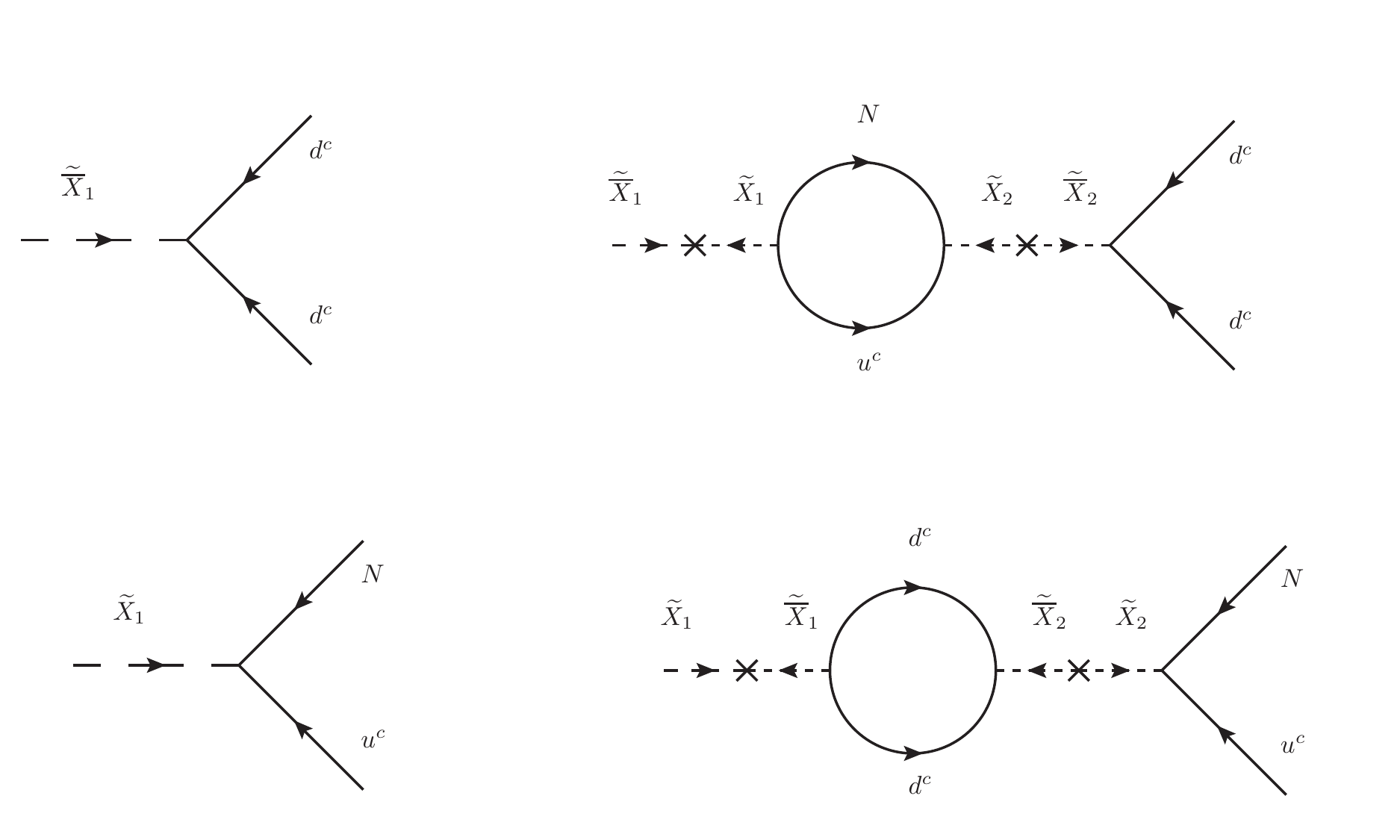}
\caption{Tree-level and self-energy diagrams responsible for generating baryon asymmetry from the decay of ${\tilde {\bar X}}_1$ and ${\tilde X}_1$. Similar diagrams for decay of ${\tilde {\bar X}}_2,~{\tilde X}_2$ are obtained by switching $1 \leftrightarrow 2$.}
\end{figure}

The way baryon asymmetry arises is quite interesting. In the limit $M_N=0$, one can assign a baryon number $+1$ to the $N$-field so that the model conserves baryon number. If we call the baryon number of quarks to be $B_q$ and that of $N$-field to be $B_N$, the total baryon number $B_{\rm tot}=B_N + B_q$ is what is conserved for $M_N=0$. Therefore, by Sakharov's criterion, the net asymmetry in $B_{\rm tot}$ produced by ${\tilde X},{\tilde {\bar X}}$ decay in this limit must vanish. As a result, any asymmetry in $B_q$ is balanced by the asymmetry in $B_N$ keeping the net asymmetry in $B_{\rm tot}$ zero. However, for $M_N=0$ the $N$ particle cannot be
observed, and hence we will observe the $B_q$ asymmetry as the baryon asymmetry of the universe. When the Majorana mass of $N$ is introduced, it will break $B_N$ by two units and due to Majorana nature, $N$-field will decay to $u^c d^c d^c$ as well as $u^{c*} d^{c*} d^{c*}$ with equal branching ratios and therefore will not add nor subtract from the $B_q$ asymmetry. In consequence, the $B_q$ asymmetry remains as the baryon asymmetry of the universe
(if $N$ were a Dirac fermion, which is not the case under consideration, the decays of $N$ and its anti-particle would have erased the $B_q$ asymmetry).
We note that the $N$-field decays well before the onset of big-bang nucleosynthesis (BBN) for the values of $M_N$, $\lambda$, and $\lambda^{\prime}$ discussed above.

To calculate the baryon asymmetry, we note that the primordial asymmetry produced per decay of ${\tilde X_1},~{\tilde {\bar X}_1}$ is given by
\bea \label{asymmetry}
\epsilon_{1} & \simeq & \frac{1}{8 \pi} ~ \frac{\sum_{i,j,k} {\rm Im} (\lambda^{1*}_{k} \lambda^2_{k} \lambda^{\prime 1*}_{ij} \lambda^{\prime 2}_{ij})} {\sum_{i,j} |\lambda^{\prime 1}_{ij}|^2 + \sum_{k} |\lambda^1_{k}|^2} ~ \frac{B_{1} B_{2} M_{1} M_{2} m^2_{{\tilde X}_1}} {(m^2_{{\tilde X}_1} - m^2_{{\tilde X}_2})^3} \, . \nonumber \\
& & \,
\eea
The asymmetry parameter for ${\tilde X}_2,~{\tilde {\bar X}_2}$ decay $\epsilon_2$ is obtained by switching $1 \leftrightarrow 2$. Here $B_{1}$ and $B_{2}$ are the $B$-term associated with the superpotential mass terms $M_{1} X_1 {\bar X}_1$ and $M_{2} X_2 {\bar X}_2$ in Eq.~(\ref{superpot}), respectively, while $m_{{\tilde X}_1}$ and $m_{{\tilde X}_2}$ denote the mass of ${\tilde X}_1,~{\tilde {\bar X}_1}$ and ${\tilde X}_2,~{\tilde {\bar X}_2}$ respectively. Superscripts $1$ and $2$ on $\lambda,~\lambda^{\prime}$ denote the couplings of superfields $X_1,~{\bar X}_1$ and $X_2,~{\bar X}_2$ respectively. We note that unlike the coupling $\lambda'_{12}$, which must be highly suppressed to meet the $p p \rightarrow K^+ K^+$ constraints, the couplings $\lambda'_{13,23}$ can be of order one. This allows one to have large asymmetry parameters $\epsilon_{1,2}$.

The observed baryon asymmetry normalized by the entropy density $s$, denoted by $\eta_B$, is obtained from above as follows:
\be \label{eta}
\eta_B \simeq \frac{1}{2} Y_S (\epsilon_1 + \epsilon_2) .
\ee
Here we have assumed that $S$ decays approximately equally to ${\tilde X}_1,~{\tilde {\bar X}_1}$ and ${\tilde X}_2,~{\tilde {\bar X}_2}$ quanta. One typically finds $\epsilon_{1,2} \sim 10^{-2}$ for natural values of couplings $\vert \lambda^{1,2}_i \vert \sim \vert \lambda^{\prime 1}_{13,23} \vert \sim 1$, $CP$ violating phases of ${\cal O}(1)$, and $m_{{\tilde X}_1} \sim m_{{\tilde X}_2} \sim M_{X,1} \sim M_{X,2} \sim B_{X_1} \sim B_{X_2}$. Entropy generated in the reheating process dilutes this asymmetry by the factor $Y_S$ which as discussed before is $\sim 10^{-8}$, thus giving the observed baryon asymmetry in the right range.

The ratio of DM abundance to baryon asymmetry follows from Eqs.~(\ref{DMab},\ref{eta})
\be \label{coincidence}
\frac{\rho_{{\tilde N}_1}}{\rho_B} \simeq \frac{2{\rm Br}_{{\tilde N}_1}} {\epsilon_1 + \epsilon_2} \frac{m_{{\tilde N}_1}}{m_p} .
\ee
Considering that ${\rm Br}_{{\tilde N}_1} \sim 10^{-3}$, the predicted value for $\rho_{{\tilde N}_1}/\rho_B$ can easily come in the ballpark of the observed value $\sim 6$. The model can therefore provide a natural explanation of the DM-baryon coincidence problem.

Finally we note that breaking of the $Z_{18}$ symmetry by the VEV of $\delta$-field will lead to domain walls. However, the entropy generation during $S$ decay will also dilute the contribution of the domain walls to the energy density of the universe. Furthermore, if there are Planck suppressed terms that break the $Z_{18}$ symmetry, they will be sufficient to destabilize the walls making them cosmologically safe \cite{rai}.
\vskip 2mm
\noindent
{\it {\bf Conclusion-}} We have discussed an extension of MSSM where tantalizing hints for light DM indicated by several direct detection experiments, including most recently the CDMS experiment, can be explained if the universe experiences a phase where its energy density is dominated by a late-decaying heavy scalar (like a modulus field) whose decay reheats the universe and yields the usual radiation dominated phase. The decay of this heavy field produces both the DM relic abundance as well as the baryon asymmetry which are comparable in their magnitude thus explaining the coincidence problem. The dark matter in our case is a scalar boson. A key ingredient of this model is the existence of new TeV scale colored particles which can be searched for at the LHC.
\vskip 2mm
\noindent
{\it {\bf Acknowledgement-}} The works of B.D. and K.S. are supported by DE-FG02-95ER40917.
The work of R.N.M is supported by the National Science Foundation grant number PHY-0968854. We would like to thank Rupak Mahapatra for valuable discussions


\end{document}